\documentclass[twocolumn]{aastex62} 
\usepackage[utf8]{inputenc}

\usepackage{natbib}
\usepackage{graphicx}
\usepackage{enumitem}
\newcommand{\numsim}{498}
\usepackage{verbatim}
\graphicspath{ {./images/} }
\usepackage{amsmath} 
\usepackage{mathtools}

\newcommand{\be}{\begin{equation}}
\newcommand{\ee}{\end{equation}}

\shorttitle{}
\shortauthors{}

\begin{document}

\title{Three-Dimensional Orbital Architectures and Detectability of Adjacent Companions to Hot Jupiters} 

\author{Thomas MacLean}
\affiliation{California Institute of Technology, Pasadena, CA, 91125}
\author[0000-0002-7733-4522]{Juliette Becker}
\affiliation{Department of Astronomy, University of Wisconsin-Madison, 475 N. Charter Street, Madison, WI 53706, US}

\begin{abstract}
The orbital properties of the (as-yet) small population of hot Jupiters with nearby planetary companions provide valuable constraints on the past migration processes of these systems. In this work, we explore the likelihood that dynamical perturbations could cause nearby inner or outer companions to hot Jupiter to leave the transiting plane, potentially leaving these companions undetected despite their presence at formation. Using a combination of analytical and numerical models, we examine the effects of stellar evolution on hot Jupiter systems with nearby companions and identify several possible outcomes. We find that while inner companions are generally unlikely to leave the transiting plane, outer companions are more prone to decoupling from the hot Jupiter and becoming non-transiting, depending on the system's initial orbital architecture. Additionally, we observe a range of dynamical behaviors, including overall stability, inclination excitation, and, in some cases, instability leading to the ejection or collision of planets. We also show that the effect of stellar obliquity (with respect to the mean planet of the planets) is to amplify these effects and potentially cause outer companions to attain non-mutually-transiting configurations more often. Our results highlight the complex dynamical pathways shaping the architectures of hot Jupiter systems.
\end{abstract}

\section{Introduction}
\label{sec:intro}
Our ability to classify mechanisms in planet formation has substantially increased in the recent decades as the Kepler, K2, and TESS missions have provided thousands of exoplanet systems \citep{Borucki2010, Howell2014, Ricker2015}, no two identical and each the result of a different superposition of formation pathways. 
One particular sub-population where significant progress has been made is that of hot Jupiters. These gas giants, akin to our own Jupiter but residing in much closer orbits to their host stars, play a pivotal role in influencing the dynamics of the systems they inhabit, impacting both the protoplanetary disk and the interactions between planets. 

Currently, three primary formation pathways for hot Jupiters have been proposed: tidal migration \citep{Fabrycky2007}, in situ formation \citep{Batygin2016, Boley2016}, and disk migration \citep{Lin1996}. While each mechanism offers a plausible explanation for some subset of system geometries, the specific conditions and processes that dictate the prevalence of one mechanism over another in different systems remain partially uncertain. Work by \citet{Petrovich2015} found that Lidov-Kozai migration (a subset of tidal migration) could explain the orbital architectures of up to about 20\% of hot Jupiter systems. \citet{Bailey2018} found the period distribution of the hot Jupiter population is consistent with being sculpted by in situ formation. 
More recently, work by \citet{Zink2023} has indicated that the vast majority of hot Jupiters form via tidal migration (specifically, via the coplanar high-eccentricity migration pathway as suggested by \citealt{Petrovich2015b}). 

Feeding into these constraints are the demographics of the hot Jupiter sample itself. 
Early results from Kepler \citep{Steffen2012} reported that hot Jupiters had a paucity of nearby {($<$1 AU)} planetary companions. Their apparent loneliness was taken as evidence for the tidal migration pathway, as hot Jupiters migrating via tidal migration would destabilize any nearby companions and explain the lack of such companions that was seen. However, that paradigm was shortly subverted by subsequent discoveries of systems where hot Jupiters coexist with {adjacent (those with orbital periods less than 30 days)} planetary companions \citep{Becker2015, Canas2019,Huang2020, Hord2022, Sha2023, Maciejewski2023, Korth_2024}. These discoveries challenged the earlier belief that hot Jupiters were solitary giants within their respective systems. The existence of nearby companions in these systems implies that at least some hot Jupiters form through less violent processes, such as \emph{in situ} formation or disk migration.

While the number of systems containing hot Jupiters and adjacent planets is small still, the discoveries over the past few years have started to reveal some trends in architectures. 
Interestingly, while only one known system (WASP-47) hosts an adjacent exterior companion, all known systems with hot Jupiters and nearby companions feature inner short-period companions (see Figure 1). Some of these companions are ultra-short-period planets (USP planets; defined as those with orbital period of less than a day or so). These USP planets, constituting about 0.5\% of all planets (\citealt{Sanchis-Ojeda2014}; though likely a larger fraction in tightly-packed, multi-planet systems; \citealt{Adams2021}), are situated in regions where stellar dynamics, particularly the slow stellar spin down observed as stars age, can significantly influence their orbits \citep{Li2020, Becker2020, Brefka2021, Schultz2021, Chen2022, Faridani2023, Faridani2025}.

\begin{figure}
\centering
\includegraphics[width=0.45\textwidth]{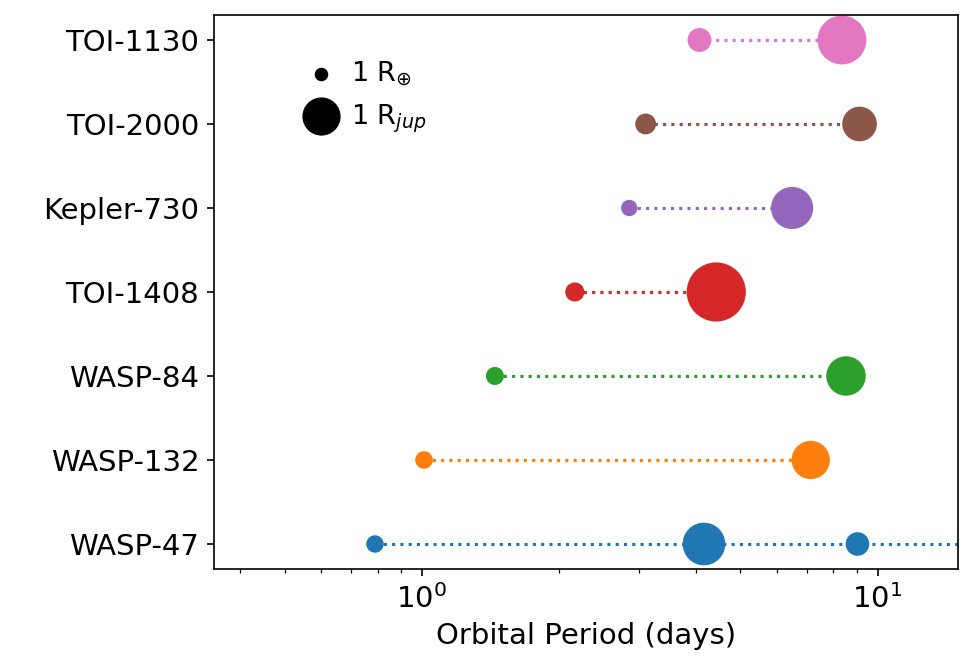}
 \caption{A schematic showing the architectures of the known systems hosting a hot Jupiter and adjacent companion planets. In these the six known systems, only the hot Jupiter in the WASP-47 system possesses both an inner and an outer companion, while the remainder of the known systems are seen to host only inner companions. }
    \label{fig:arch}
\end{figure}

In this work, we consider the dynamics of exoplanet systems containing hot Jupiters with adjacent planetary companions, particularly in the presence of evolving host stars. 
Using a combination of analytical and numerical calculations, we model the evolving detectability of these companions against the backdrop of the spin-down of their host stars, assessing the likelihood that initially coplanar companions of hot Jupiters remain observable in multi-planet configurations today.
The paper is structured as follows: first, Section \ref{sec:sec2} presents an analytical model that lays the foundation for understanding the problem at hand.
In Section \ref{sec:sec3}, we turn to numerical simulations to model the long-term evolution of the hot Jupiter-hosting systems as their host star ages. 
In Section \ref{sec:sec4}, we apply our modelling framework to the population of observed systems that match this architecture, with an emphasis on determining how stellar obliquity affects the observed dynamics.  
In Section \ref{sec:disc}, we explore the broader implications of our findings, including their relevance to the dynamics and detectability of hot Jupiter systems. Finally, in Section \ref{sec:conclude}, we provide a summary of our conclusions and highlight key takeaways from this work.

\section{Theoretical Motivation}
\label{sec:sec2}
In this work, we aim to determine plausible dynamical outcomes for hot Jupiter systems with adjacent inner and outer companions post-disk dissipation. To build a physical intuition for the problem, we start by employing a secular approach to modeling the planetary evolution. In this model, we include planet-planet interactions as well as the spin-down of the star, modeled by a time-evolving stellar quadrupole moment ($J_2$). 


We begin by defining the secular perturbation function up to second order terms, following the convention of \citet{MD99}. We exclude short-period terms which average to zero over long timescales. The evolutions of planetary inclination and eccentricity are conjugate yet decoupled in the second-order secular theory, so we consider the disturbing function as a function of only inclination:
\be
\mathcal{R}^{\rm{(sec)}}_{j} = n_{j}a_{j}^{2} 
\Biggl[ \frac{1}{2} B_{jj} I_{j}^{2} 
+\sum_{k=1}^{N} \left( B_{jk} I_{j}I_{k} \cos{(\Omega_{j} -\Omega_{k})}\right)   
\Biggr]\,,
\ee
where $I$ is the planet inclination variable, $n$ the mean anomaly, $\omega$ the argument of pericenter, $\Omega$ the longitude of ascending node, $j$ the index of the planet in the system, $k$ the index of the perturbing body, and $B_{ij}$ the frequency coefficients which form the \textbf{B} matrix \citep{MD99}. 
To express these matrix coefficients, we require further definitions which rely on the semi-major axes of the planets. We first define $\alpha_{jk}$ to be in the interval $(0,1)$; this quantity represents the semi-major axis ratio of planet j over planet k when the semi-major axis of planet k is larger than that of planet j. We define the quantity $\bar{\alpha}_{jk}$ to be 1 when the semi-major axis of planet j is longer than that of planet k (representing an internal perturber) and otherwise it is defined the same as $\alpha_{jk}$ \citep{MD99}. The cases where these quantities are defined the same denote external perturbers. 

The coefficients $B_{ij}$ are defined in terms of only masses and planet semi-major axes: 
\be
B_{jk} = n_{j} \left[ \frac{1}{4}\frac{m_{k}}{M_{c} + m_{j}}
  \alpha_{jk} \bar{\alpha}_{jk} b^{(1)}_{3/2}(\alpha_{jk}) \right]\,,
\label{bmatrixoff} 
\ee 
and
$$
B_{jj} = -n_{j} \Biggl[ \frac{3}{2} J_{2} 
\left( \frac{R_{c}}{a_{j}} \right)^{2} - \frac{27}{8} J_{2}^{2} 
\left( \frac{R_{c}}{a_{j}} \right)^{4} 
$$
\be
\qquad \qquad  + \frac{1}{4} \sum \frac{m_{k}}{M_{c} + m_{j}}
\alpha_{jk} \bar{\alpha}_{jk} b^{(1)}_{3/2}(\alpha_{jk}) \Biggr] \,,
\label{bmatrixdiag} 
\ee Again we let $j$ and $k$ represent planet indices, with the diagonal terms indicative of the interactions between the planet indexed by $j$ and the central star. 
The term $b^{(1)}_{3/2}$ is a convergent Laplace coefficient, whose value can be found in \citep{MD99}. The $m$ represents the mass of each planet, and $M_c$ the mass of the central body. The term $b^{(1)}_{3/2}$ is a convergent Laplace coefficient. The $m$ represents the mass of each planet, and $M_c$ the mass of the central body. 
The \textbf{B} matrix defines a compact notation for the inclination evolution of the planets. For the diagonal terms of \textbf{B}, we also consider the star-planet interactions, quantified by the stellar oblateness $J_2$. 
We choose a physical way to define $J_2$, as in \citet{Ward1976}:
\be 
J_{2} = \frac{k_{2}}{3}\left[\frac{\Omega_{\textit{star}}}{\Omega_{\textit{breakup}}} \right]^2 \,,
\label{J_2} 
\ee 
where the Love number is $k_2$, stellar rotation frequency $\Omega_{\textit{star}}$, and frequency at break-up of the star $\Omega_{\textit{breakup}}$. 
As stars spin down, their $J_2$ will decrease with time. 
As seen in Equation \ref{bmatrixdiag}, the impact on the disturbing function of $J_2$ falls off with $a^2$. To construct the diagonal terms of the \textbf{B} matrix (Equation \ref{bmatrixdiag}, we neglect the higher order moments ($J_4$ and above).
In systems containing hot Jupiters with interior planetary companions, both planets are often situated within ten stellar radii, putting the planets in the regime where $J_2$ dynamics will be most relevant.  

We next solve for the eigenvectors and eigenfrequencies $f_{k}$ of the \textbf{B} matrix, which allows us to write the phases $\gamma_k$ as determined by the initial conditions.  The initial conditions correspond to observations of the inclination at some time which can be used to derive the phases and scaling of eigenvectors \citep{MD99}. 
We can compactly represent the inclination evolution of each planet in time by:
\be 
\begin{split}
I_{j}(t) =& \Bigg( \left[\sum_{k=1}^N I_{jk} \sin (f_k t + \gamma_k) \right]^2 +\\
&\left[\sum_{k=1}^N I_{jk} \cos (f_k t + \gamma_k) \,,
\right]^2 \Bigg)^{1/2} \,,
\end{split}
\label{inc_excite} 
\ee 

This equation of motion allows us to compute the time-evolving planet inclination as it is affected by both planet-planet and planet-star interactions. 
Following this secular framework, \citet{Spalding2016} found a resonant criterion for two-planet systems based on the secular theory outlined above. Secular resonance will onset when:
\begin{equation}
\begin{split}
J_{2} \big|_{\text{res}} \approx \frac{1}{6} \left( \frac{a_1}{R_{\star}} \right)^2 \left( \frac{a_1}{a_2} \right)^2 \left( \frac{m_2}{M_{\star}} \right) \left( \frac{b_{3/2}^{(1)}(\alpha)}{m_2 \sqrt{G M_{*} a_2}} \right)\\
\times \frac{(m_1 \sqrt{G M_{*} a_1} + m_j \sqrt{G M_{*} a_2})^2}{(m_1 \sqrt{G M_{*} a_1} - m_2 \sqrt{G M_{*} a_2})(1 - \alpha_{12}^{7/2})},
\end{split}
\label{eq:j2res}
\end{equation}
where $1$ and $2$ subscripts denote the inner and outer planets. 
For a two planet system consisting of a inner sub-Neptune at 0.02 AU and a hot Jupiter at 0.04 AU, this resonance will not be excited for positive values of $J_2$, since the outer planet has more angular momentum than the inner one. 
However, for a two-planet system with a hot Jupiter and an exterior sub-Neptune companion at 0.09 AU, the hot Jupiter has more angular momentum and a resonance will be excited at $J_{2} \big|_{\text{res}} \approx 7\times10^{-3}$.

The secular theory suggests that hot Jupiter-hosting multi-planet systems may provide an environment where the onset of the $J_2$ secular resonance as the star spins down will play a significant role in setting the observed system architecture. In particular, as indicated by Equation \ref{eq:j2res}, we expect mutual inclinations to be excited in pairs of planets consisting of hot Jupiters and outer, smaller planets. However, to fully assess the coupled dynamics at play in these systems and evaluate how well this secular prediction holds in three planet systems, we next turn to more comprehensive N-body simulations.





\section{Numerical Results}
\label{sec:sec3}
In this section, we investigate the dynamical evolution of multi-planet systems with hot Jupiters, focusing on the impact of stellar spin-down. For these simulations, we set up the planetary geometry using exemplar system WASP-47. 
Among the planetary census, the WASP-47 system stands unique in hosting a hot Jupiter flanked by both interior and exterior nearby planetary companions. This configuration has not yet been observed in any other known planetary system. The other six published systems with hot Jupiters and nearby companion planets feature only inner companions. Therefore, WASP-47 represents the most complex known arrangement of a hot Jupiter with the greatest number of close planetary neighbors. 

The relevant orbital and physical parameters of the planets in the WASP-47 system are given in Table \ref{tab:hot_jupiter_systems}. 
We note that WASP-47 also has, in addition to the two nearby companions, a massive, distant companion \citep[a Jupiter-mass planet at an orbital period of $\sim$590 days;][]{NeveuVanMalle2016}. For the purposes of this work, we consider this companion dynamically decoupled and assume its orbit is oriented roughly coplanar to the inner system such that it does not significantly precess the plane of the inner planets with respect to the host star \citep{2017Vanderburg, Becker2017}.

Our initial aim in this section is to create realizations of the WASP-47 system by adjusting its parameters and examine how these modifications affect the system's detectability. In this section, we call the innermost planet in all realizations of the WASP-47 system the `USP planet', the hot Jupiter WASP-47b the `hot Jupiter', and the outer companion WASP-47d the `outer planet'. The relative orderings of their separations is not changed in any simulation (the order from closest to farthest from star is always USP planet, hot Jupiter, outer planet). 
 \begin{figure}
\centering
\includegraphics[width=0.45\textwidth]{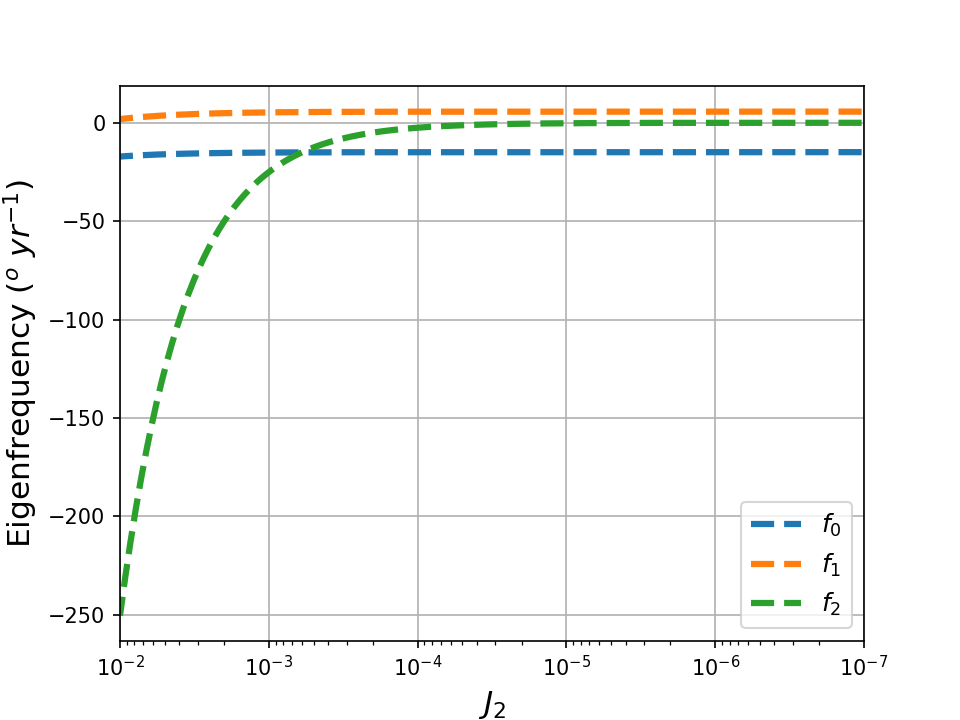}
\caption{Plot of eigenfrequencies for each of the modes in the three-planet WASP-47 system as determined by the secular model described in Section \ref{sec:sec2}. Which mode dominates the dynamics of the system will depend on the value of $J_2$. }
    \label{fig:eigenfreq}
\end{figure}
\subsection{Numerical Simulations: Setup}

This study utilizes the Rebound simulation package, incorporating Reboundx for general relativity (GR) and $J_2$ effects \citep{Tamayo_2019, Tamayo2020, Tamayo2023, Rein_2012}. For our initial analysis, we assume a fixed stellar obliquity $\lambda$ with respect to the orbital plane of the hot Jupiter throughout our simulations to examine the isolated impact of the precession induced by star’s gravitational moment ($J_2$, which induces precession in $\Omega$ and decreases with time as the star spins down) and GR (which induces precession in $\omega$) on the system. 

In our initial suite of simulations, we focus on the effects of altering the semi-major axes of the hot Jupiter and its adjacent planetary companions. The two primary variables under consideration in this analysis are the semi-major axis of the innermost planet ($a_{USP}$) and the semi-major axis ratios of both pairs of planets ($\alpha_{USP-HJ}$, $\alpha_{HJ-OUT}$). These are chosen as the most important variables because the orbital distance of the innermost planet, combined with the ratios of semi-major axis for all planets in the system, directly set the importance of the precession due to the host star's quadrupole moment as compared to the importance of planet-planet secular interactions. 

Using Rebound's \texttt{ias15} integrator, we performed simulations over durations of \(10^5\) years. 
Since this work considers a system's behavior in relation to the star's $J_2$, we decrease $J_2$ logarithmically between $J_2 = 10^{-3}-10^{-8}$ over the course of the $10^5$-year integrations.
This integration duration was selected to capture the relevant dynamical changes with $J_2$ while maintaining computational efficiency. To validate this choice, we conducted a test suite of simulations with identical input parameters but varying integration lengths. For integration lengths of \(10^5\) years and beyond, no differences in dynamical behavior were observed, indicating that this timescale accurately reproduces the system's evolution. 
{For additional confirmation of the appropriateness of this integration time, we use the best-fit values from Table \ref{tab:hot_jupiter_systems} and second-order Laplace-Lagrange theory to find that the largest eigenvalue of the B matrix in the WASP-47 system is 1.4$\times10^{-4}$, corresponding to a secular timescale of approximately 9500 years. This timescale is significantly shorter than our chosen integration time, meaning that our integrations are sufficiently long to capture this evolution. }

We run a total of \numsim\ simulations of WASP-47 analogues (including planets e, b, and d, but excluding dynamically decoupled companion WASP-47 c), with orbital parameters set by default to the best-fit values given in Table \ref{tab:hot_jupiter_systems}. For all simulations, we set the {range of the} initial orbital inclinations of each planet to be {one degree, so that the orbits of all three planets were roughly coplanar. Following} the observations of WASP-47, we set all planetary eccentricities $e=0$, we set stellar mass to be 1 solar mass, we set stellar radius to be 1 solar radius, and we set the obliquity of the star with respect to the average plane of the planets ($\lambda$ = 10 degrees).

To explore the effect of orbital spacing on the system evolution, we scale the observed semi-major axis of each planet by a factor $\zeta$, such that  $a_{USP} = \zeta_{USP} a_{USP,0}$ denotes the semi-major axis of our simulated realization of the ultra-short-period planet (the true measured semi-major axis of which is $a_{USP,0}$).
Analogously, we also define $\zeta_{HJ}$ for the Hot Jupiter (planet b). 
In each of the \numsim\ simulations, $\zeta_{USP}$ and $\zeta_{HJ}$ are selected randomly from uniform distributions $\zeta_{USP} \sim \operatorname{Uniform}(0.5, 3)
$ and $\zeta_{HJ} \sim \operatorname{Uniform}(a_{USP,0}, 1.4)$ under the constraint $a_{USP}< a_{HJ}$.

Our \numsim\ simulations have drawn from the bounds on $\zeta_{USP}$ and $\zeta_{HJ}$ described above, while the orbital radius of the outer planet is not changed and instead set to the best-fit value in Table \ref{tab:hot_jupiter_systems}.

\begin{figure*}
\centering
 \includegraphics[width=1\textwidth]{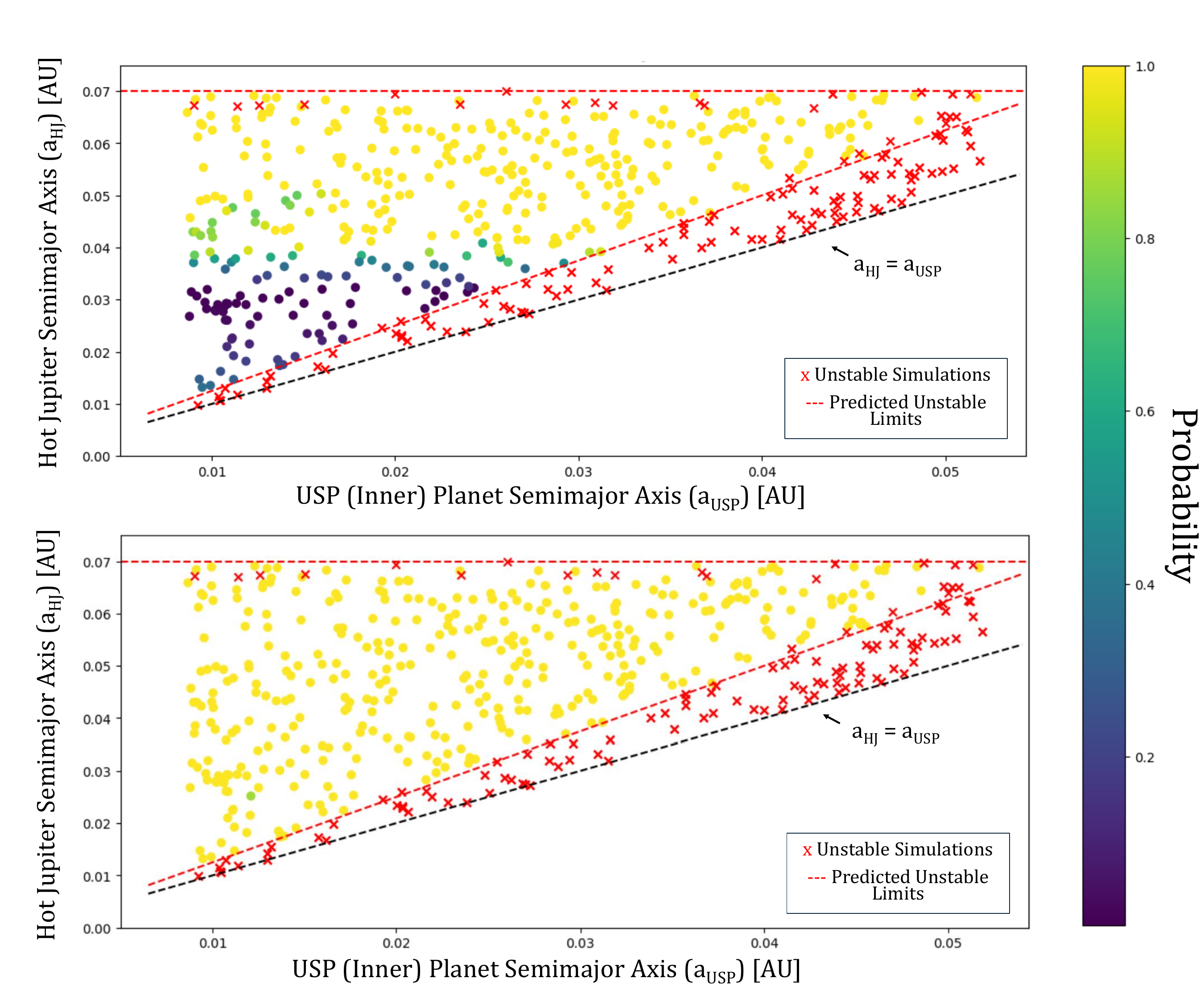}
 \caption{Phase space plot of the (top panel) outer planet's transit probability and (bottom panel) inner planet's transit probability over the range of planetary parameters tested in the suite of simulations. To calculate the transit probability, we compute the impact parameter as defined by the condition from Equation \ref{equation:trthresh} for each time-step in each simulation, and we calculate the transit probability by averaging the state of the system over the last 10 percent of the integration. The pair $(a_{USP}, a_{HJ}) = (0.0173, 0.05)$, which resides in a region of the plot with high transit probability, corresponds to initial conditions matching the currently observed state of WASP-47. Dynamically unstable integrations are denoted by red 'X' markers.} 
    \label{fig:phases}
\end{figure*}

\subsection{Numerical Simulations: Results}
Variations in the dynamical evolution across our cases arise from the specific values of $a_{USP}$ and $a_{HJ}$ selected for each simulation.
Additionally, as discussed in Section \ref{sec:sec2}, the variations in mutual inclination between adjacent planets are caused by the system passing through a secular resonance as $J_2$ decreases. 
As a motivating example, in Figure \ref{fig:eigenfreq}, we show how the eigenfrequencies of matrix \textbf{B} change for the exemplar best-fit WASP-47 system.

Our numerical simulations allow us to resolve the full behavior in the three-planet system as it passes through secular resonance.
As we are changing the spacings of the planets in the system between our \numsim\ simulations, the exact value of $J_{2}$ at which the secular resonance will occur will change (in a way modeled by Equation \ref{eq:j2res}). 

Our primary motivation in running these simulations was to assess how an observer's impression of the number of planets in a three-planet system containing a hot Jupiter (like WASP-47) depends on the planet parameters $a_{USP}$ and $a_{HJ}$. 
Our ability to recover the presence of additional planets in hot Jupiter systems has so far required transit detection of the additional companions, which requires those planets to be mutually transiting with the hot Jupiter.

\begin{figure*}
\centering
 \includegraphics[width=\textwidth]{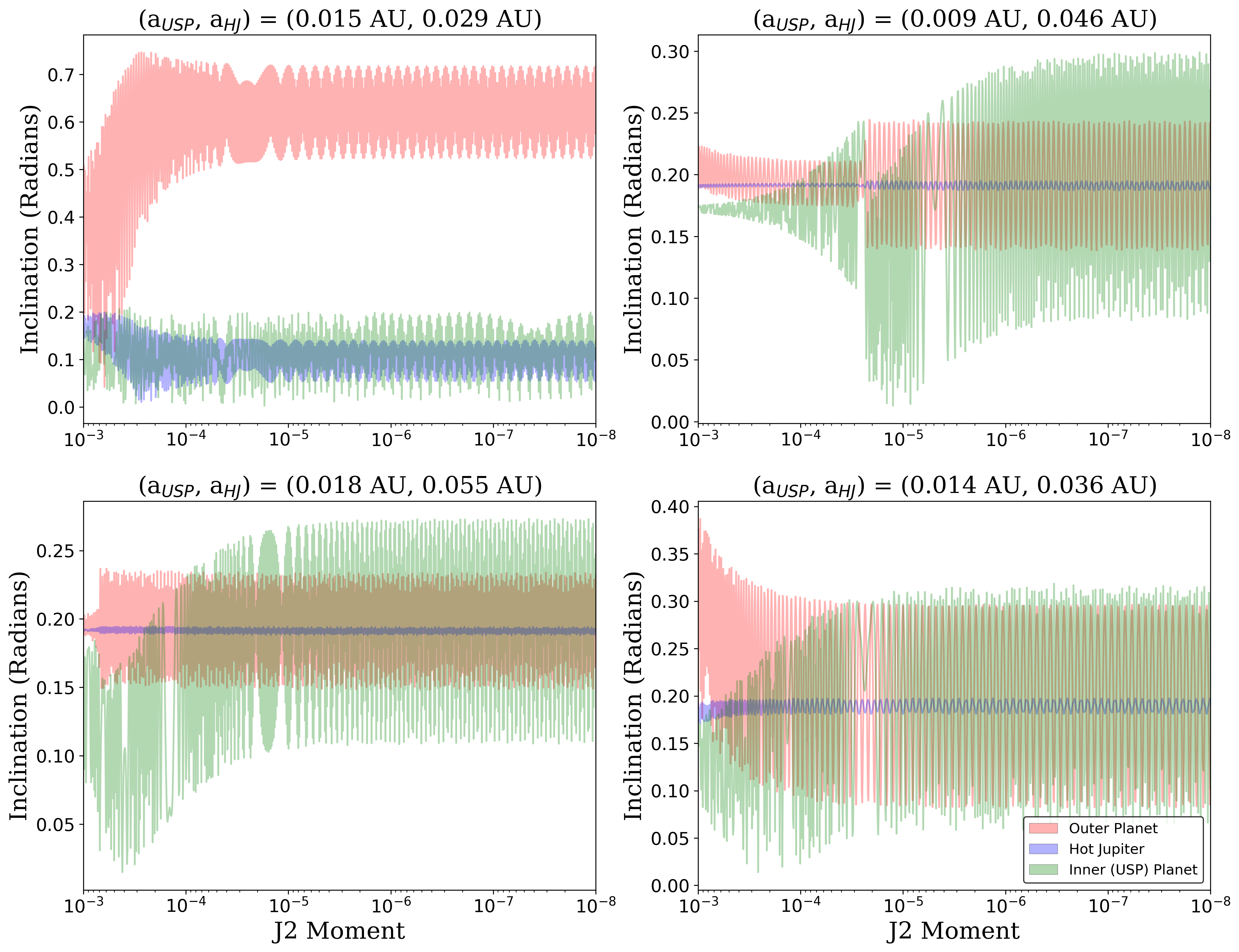}
 \caption{Our simulations show a range of dynamical behavior resulting from planet-planet interactions and the evolving effect of the stellar quadrupole $J_2$. There are several notable regimes of behavior: (upper left panel) dynamical stability where the outermost planet's inclination angle becomes decoupled from that of the other two planets; (lower left panel) a dynamically stable configuration where all planets retain similar inclination angles; (lower right panel) dynamical stability where inclination amplitudes are excited sufficiently for the outer planet such that not all planets can generally be seen to transit at once; (upper right panel) dynamical stability where the innermost planet experiences a relative increase in inclination amplitude compared to the other planets, yet remains transiting; (not depicted in the figure) and finally, dynamical instability where at least one planet collided with another body or was lost from the system.}
    \label{fig:modesplot}
\end{figure*}

Our simulations show several distinct dynamical regimes, defined by how they will be seen in transit. 
The distinct dynamical regimes have different implications with regards to which planets can be seen in transit simultaneously with the Hot Jupiter. 
We show representative plots of the four most common evolution pathways in Figure \ref{fig:modesplot}. These pathways are as follows: \vspace{-2mm}
\begin{enumerate}[noitemsep]
    \item \emph{Decoupling}: At low \(\alpha_{HJ-OUT}\), the outer companion's inclination decouples from the inner planets due to stellar \(J_2\)-induced precession, leading to significant mutual inclination changes and AMD redistribution (Figure \ref{fig:modesplot}, upper right). Similar behavior can be seen be seen in Figure 4c of \citet{Chen2022} and Figure 4 of \citet{Brefka2021}. This effect was also discussed in \citet{Spalding2016} and \citet{Batygin2016}. This behavior is shown in the upper right panel of Figure \ref{fig:modesplot}. In this geometry, the inner planet could be seen in transit at the same time as the hot Jupiter, but the outer planet could not. 
\item \emph{Stable and Transiting}: In this configuration, all planets remain dynamically stable with small inclination variations, ensuring they continue to transit. This geometry can be seen for parameters similar to those seen the in true WASP-47 system. This behavior is shown in the lower left panel of Figure \ref{fig:modesplot}. In this geometry, the all three planets would be discoverable via transits from the same line of sight.

\item \emph{Oscillation-Driven Non-Transiting Configurations}: In this configuration, the planets all remain in dynamically stable orbits centered around the same orbital planet, but the oscillation amplitudes their orbital inclinations become large enough to lead to one of the planets (or more) to not be simultaneously transiting with the others.  This behavior is shown in the lower right panel of Figure \ref{fig:modesplot}.

\item \emph{Dynamical Instability}: For high \(\alpha_{USP-HJ}\) values (\(>0.77\)), planets' orbits evolve over time in semi-major axis, leading to system destabilization, which we define as ejections or collisions between any bodies in the system (planets and star). Although not depicted in Figure \ref{fig:modesplot}, such systems may exhibit similar behavior to the panels of Figure \ref{fig:modesplot} up until instability; these systems are shown as red X's in Figure \ref{fig:phases}. 

\end{enumerate}

Each of these modes of behaviors leads to a different subset of the system planets being detectable in transit (from any line of sight). 
The criterion for a planet to be transiting is computed as follows. 
The geometry of a planetary system relative to the observer yields a threshold for transiting, which is dependent on the planetary inclination ($i$, where $i=90$ degrees denotes an orbit seen edge-on), object radii, and planet semi-major axis $a$ and is given by:
\begin{equation}
    \cos{i} \le \frac{R_p + R_*}{a}
    \label{equation:trthresh}
\end{equation} 
Assuming that $R_p << R_*$, we can rearrange this threshold for transiting in terms of the impact parameter $$b = \frac{a \cos{i}}{R_*},$$ such that planets will transit if their impact parameters are $-1 < b < 1$. We compute this impact parameter $b$ for each planet at each time-step in our simulations, assuming a line of sight that maximizes the transit probability conditional on the hot Jupiter being seen in transit. To find the final transit probability for a configuration, we compute the transit probability of each planet pair over the last 10\% of the integration for each of our \numsim\ simulations.

These computed transit probabilities are plotted in Figure \ref{fig:phases}. Each individual simulation generates one point in each of the two panels of Figure \ref{fig:phases}: in the top panel, we show the transit probability of the outer planet given that the Hot Jupiter transits; in the bottom panel, we show the transit probability of the ultra-short-period planet given that the Hot Jupiter transits. We note that the outer planet's semi-major axis was not altered for the purposes of this numerical experiment, which is why the plot axis show only the varied values of $a_{HJ}$ and $\mathbf{a_{USP}}$.

We note that the bounds of the parameter space of our simulations include orbits beyond the expected dynamical limit of stability where the planets' Hill spheres overlap. 
The mutual Hill radius between two planets is defined as \citep{chambers1996}:
\begin{equation}
    R_{Hill,12} = \left(\frac{M_{1}+M_{2}}{3M_{\odot}}\right)^{1/3}\left(\frac{a_{1}+a_{2}}{2}\right),
    \label{rhill}
\end{equation}
In Figure \ref{fig:phases}, we plot the two stability limits as dashed lines: (1) in red, the dashed lines indicate regions where two planets in the system fall within $2\sqrt{3}R_{Hill}$ of each other, which suggests dynamical instability \cite{chambers1996, Marzari2014}); (2) in black, we plot the boundary at which $a_{USP} = a_{HJ}$. 

The upper panel of Figure \ref{fig:phases} reveals a wide range of transit probabilities for the outer companion in the systems simulated, indicating varying levels of observability across the simulations. These results appear to indicate that the transiting probabilities are dependent on the hot Jupiter orbital period, and some parameter combinations lead to the outer planet being outside of the transiting plane, which matches our predictions from Section \ref{sec:sec2} that the resonance of the hot Jupiter could in some cases affect the inclination evolution of the outer planet. 

The lower panel of Figure \ref{fig:phases} reveals that the innermost planet remains largely transiting across our suite of simulations, which indicates that the USP (innermost planet) whenever the hot Jupiter transits given that we initialize the simulations with the hot Jupiter assumed to be transiting and the inner USP planet in the same plane. This indicates a high probability that the inner planet will be transiting, given one exists, given that the hot Jupiter is viewable in the transiting plane. This matches current observations, as well as predictions from Equation \ref{eq:j2res} in Section \ref{sec:sec2} that the USP planet will not cross the resonance and is likely not to de-couple in stable system arrangements.

Our results show that the USP planet consistently remains transiting when the hot Jupiter is observed in transit. This aligns with our initial setup, where the USP planet and hot Jupiter began in the same orbital plane, and with the predictions from Section \ref{sec:sec2}, particularly Equation \ref{eq:j2res}, which predicts that the USP planet should not cross the secular resonance due to having less angular momentum than its adjacent hot Jupiter companion. 
In contrast, the outer planet’s transit behavior varies across parameter combinations, with some configurations leading to it leaving the transiting plane. 
This behavior highlights the role of dynamical interactions and secular resonance crossings in determining the visibility of outer companions.

\section{Hot Jupiter Systems with Nearby Companions}
\label{sec:sec4}
In the previous section, we examined how a single exemplar system (WASP-47)'s observability may have been affected by slightly different orbital geometries (in terms of the planetary semi-major axis ratios $\alpha$).  
As demonstrated by the results shown in Figure \ref{fig:phases}, while the USP-hot Jupiter planet pair never attain a mutually non-transiting configuration (which Equation \ref{eq:j2res} suggests may be because they never cross the secular resonance as $J_2$ decays), the hot Jupiter-outer planet pair do cross the resonance and sometimes attain mutually non-transiting configurations (particularly for lower values of $a_{HJ}$). 
Example trajectory plots (see the top left and bottom right panels in Figure \ref{fig:modesplot}) demonstrate that this can lead to the outer planet not being seen in transit from the plane of the observer where the hot Jupiter can be seen. 

The census of hot Jupiters with nearby planetary companions (Figure \ref{fig:arch}) shows that WASP-47 is presently an outlier - the only system to be observed to have a nearby outer companion but not a nearby inner companion. 
The lack of an outer companion could occur for several reasons: some classes of hot Jupiter formation models \citep{Batygin2016} predict this lack of outer planets, geometrical transit probabilities are lower for outer companions, and it is also possible that planets may exist in transiting configurations but be too small to be discovered. 

While in Section \ref{sec:sec3} we fix the stellar obliquity to 10 degrees to reduce the number of degrees of freedom in our simulation suite, it is likely that these stellar hosts could have a range of initial stellar obliquities with respect to the plane containing the planets. While evidence in the literature \citep{Morton2014, Becker2017} suggests that stellar obliquities of cool stars hosting multiple planets (including hot Jupiter multi-planet systems) are low, theoretical expectations \citep[i.e.,][]{Batygin2013, Spalding2015, Epstein2022} predict that the exact value of the initial star-disk angle may vary. 
For that reason, in this section we examine how the initial stellar obliquity would affect the transit probability of outer companions to the six hot Jupiter systems shown in Figure \ref{fig:arch}. 

\subsection{Case Studies of Known Systems}
{We next proceed by examining how an evolving stellar oblateness would affect other hot Jupiter systems with nearby companions. }
{To select systems that fit this definition, we apply several cuts to the sample. First, we require that to be defined as a `hot Jupiter', a planet must have an orbital period of 10 days or less \citep{Wang2015, Howe2025} and a mass of at least 0.25 $M_{jup}$ or, if a mass for the planet has not been measured, a radius of at least 0.75 $R_{jup}$. Second, we define adjacent companions to include those with orbital periods of less than 30 days.}

{Under these constraints, we} note that there are only 7 hot Jupiter systems that have {adjacent} planetary companions confirmed as of this paper's completion, and WASP-47 is the only confirmed to have a nearby outer companion \citep{Becker2015}. See Table \ref{tab:hot_jupiter_systems} for the full list of confirmed planetary systems with their orbital parameters. The remaining six systems (TOI-2000, Kepler-730, WASP-84, WASP-132, TOI-1408, and TOI-1130) are observed to have only inner companions. In this section, we consider how initial stellar obliquity may affect the observability of unseen nearby outer companions (analogous to WASP-47 d) in these systems. 

For each of our six observed systems, we prepare model planetary systems which are assumed to have a roughly coplanar geometry upon disk dissipation, with a hot Jupiter, nearby inner companion, and a `test' nearby outer companion. 
These six model systems are first initiated as two planet systems, with orbital periods, stellar properties, planet masses and radii, and angular orientations equal to the best-fit values in the known systems ({see} Table \ref{tab:hot_jupiter_systems}). 
The initial stellar obliquity, defined relative to the mean planet of the planets, is set to $\lambda = 1$ degree. 
From these base two-planet systems, we add a third planet exterior to the hot Jupiter with mass of 10\% percent of the hot Jupiter, zero eccentricity, and zero inclination, but an orbital distance that is allowed to vary between simulations. 
For each system, we run a number of simulations; these simulations are determined by taking a uniform selection $\{\zeta_{i}\}_i$ of 15 factors between a lower-limit factor of 1.25 and upper-limit factor of 3.1 from the hot Jupiter for the semimajor axis of the `test' outer planet. In each system, the initial parameters of the hot Jupiter and the innermost planet are held the same. In the TOI-2000 system we also run a densification algorithm to add additional simulations in regions where the difference between adjacent sampled points result in probabilities that are sufficiently large to garner interest in the behavior of the unsampled gap region. This densification algorithm is run twice, where each time that it runs, whenever `adjacent' simulations with factors $\zeta_{i}, \zeta_{i+1}$ result in outer planet transit probabilities that differ by at least 0.25, an additional simulation is run with the outer planet placed at a semimajor axis $\zeta_{j}*a_{HJ}$ where $\zeta_{j} = \frac{\zeta_{i}+\zeta_{i+1}}{2}$. 
In each simulation, in an identical procedure to that used in Section \ref{sec:sec3}, the stellar $J_2$ is allowed to decrease logarithmically between $10^{-3} - 10^{-8}$ over an integration time of $10^5$ years integrated with \texttt{ias15}, and the effects of GR are included via the Reboundx implementation.  
Finally, we repeat the process outlined above for four other values of initial stellar obliquity, resulting in a simulation set for each of the six systems with results for $\lambda = [1, 5, 10, 20, 30]$ degrees.

\begin{figure*}
\centering
 \includegraphics[width=\textwidth]{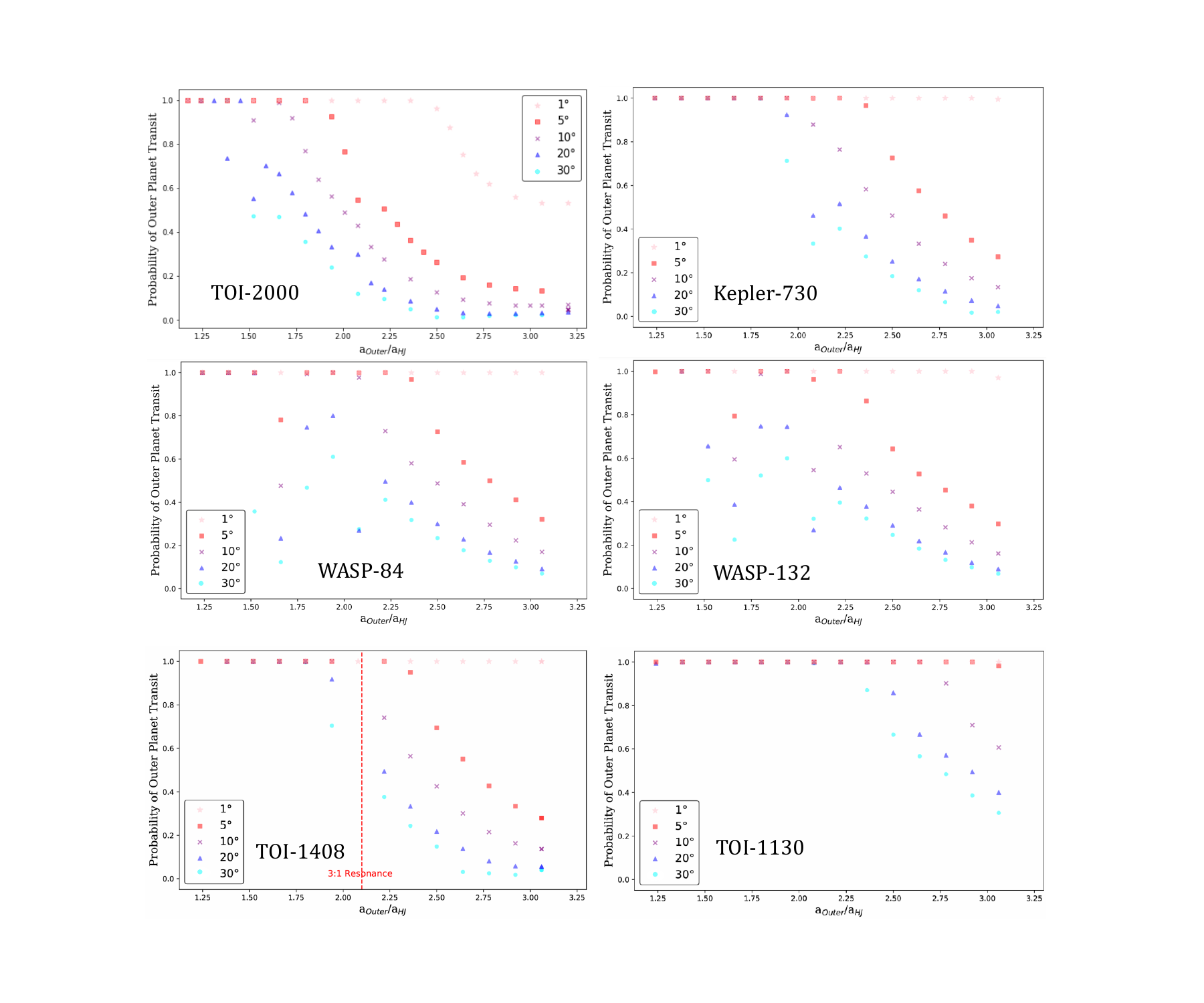}
 \caption{The six systems depicted are all those with inner companions to hot Jupiters. In each system, we simulate the planetary evolution by running simulations where each system begins with a ``test" outer planet adjacent to the hot Jupiter in the transiting plane. Each plot shows the probability of transit for the outer planet versus the semi-major axis ratio of the outer planet after the full time integration, with each point representing a different simulation with corresponding semi-major axis ratio of the outer planet to the hot Jupiter and stellar obliquity. We find that as the outer planet moves away from the hot Jupiter and as stellar obliquity increases, in accordance with specific properties of each individual system, the probability of transit of the outer planet tends to go down, across the simulations we ran. }
    \label{fig:TOI2000}
\end{figure*}

For each simulation we run, we compute the present-day transit probability using the procedure outlined in Section \ref{sec:sec3}.
The transit probabilities computed for this set of simulations are presented in Figure \ref{fig:TOI2000}. If stellar obliquity had no effect on the transit probability of the outer planet, we would expect the transit probability curves for a single system to decrease with larger $a_{OUT}/a_{HJ}$ due to geometrical effects. 
However, we find that stellar obliquity is significant in detectability of a hot Jupiter’s nearby outer companion: for all six systems under consideration, the transit probability of the `test' outer planet is significantly higher for lower stellar obliquities. This follows the result of \citet{Spalding2016} and \citet{Brefka2021}, both of which found that inclination oscillation amplitudes are larger for larger stellar obliquities. 

Additionally, we find that across our systems tested, the nature of the transit probability ``drop-off" is varied, likely as a result of the differing system orbital parameters. For a system like TOI-1130, which has the inner planet-hot Jupiter pair the furthest from their host star (and thus a less significant contribution to the dynamics from the evolving stellar $J_2$), even large stellar obliquities ($>20$ degrees) only have minor effects on the observability of outer companions within an $\alpha = a_{OUT}/a_{HJ} <2$. On the other hand, systems like TOI-2000, which has a `hot Jupiter' with significantly less angular momentum than in the other systems, have decreased transit probabilities for a large range of $\alpha$ values. 

\vspace{-4mm}
\section{Discussion}
\label{sec:disc}

Our analysis explored the dynamical evolution of tightly-packed, multi-planet systems containing hot Jupiters after disk dissipation. 
By examining a model system with a hot Jupiter sandwiched between two nearby planetary companions (WASP-47), we demonstrated the range of interactions in this type of multi-planet system.
The dynamical outcomes in this first model system ranged from systems maintaining relative stability, with minimal inclination changes, to scenarios involving large shifts that caused decoupling or disrupted the potential for transit visibility, especially for outer companions.
These interactions generally will not excite significant mutual inclination between a hot Jupiter and an inner companion. 
On the other hand, in those same systems, it is much easier to excite mutual inclination between a hot Jupiter and its outer companion. This highlights the rich variety of evolutionary pathways for hot Jupiter systems with nearby planets.

Following that analysis, we examined case studies of the six other known systems that host hot Jupiters and at least one nearby companion planet. 
We ran simulations for each system with multiple values of stellar obliquity, examining the observability of a `test' outer planet in each system as a function of its orbital radius and the stellar obliquity. 
We found that stellar obliquity significantly impacts the transit probability of outer planets, with higher stellar obliquities reducing the likelihood of detecting the outer planet, conditional on the hot Jupiter being observed in transit. These results suggest that both the obliquity of the host star and the orbital distance of planets play key roles in determining the detectability of adjacent planets. In particular, more distant planets and those in systems with higher stellar obliquity are less likely to be detected.

Future studies could expand on this work by exploring broader system configurations, varying orbital distances, and stellar obliquities, and examining warm Jupiter systems, where the larger separations might lead to different dynamical regimes. 
{This work and previous work by \citet{Faridani2023} considered three planet systems, but more complex geometries may result in different evolutionary pathways than those considered in this work. Similarly, we focus on a fairly narrow system geometry, and the growing population of hot Jupiters with nearby companions demonstrates that there is significant diversity within the sample, including mean motion resonance \citep{Korth_2024}.}
Such efforts would deepen our understanding of the full range of possible system architectures and the mechanisms driving their evolution.

Significant misalignment of the inner companion is likely to be associated with a dynamical instability in the system, making it unlikely to occur in observed systems with our range of tested parameters. 
Therefore, it is unlikely that a perturbation of the nature considered in this work would lead to a USP inner companion residing with a high misalignment from a hot Jupiter.


{\subsection{Inferences on Planet Formation from Hot Jupiter Companion Multiplicity}}
{This paper investigates systems that host nearby companions to hot Jupiters and finds that inner companions to hot Jupiters, if they exist in dynamically stable configurations, are likely to remain mutually  transiting with the hot Jupiter.}
{This result indicates one of two possibilities in systems where inner companions are not seen: 1) that these systems may be truly isolated, undergoing distinct evolution pathways from hot Jupiter systems with nearby planets; or 2) the planets exist but are of too small radii to have been detected photometrically. Regardless, our results show that an observed inner planet is consistent with disk-driven migration \citep{DawsonJohnson2018} and in situ migration \citep{Batygin2016}, as it is not easy for stellar-rotation-driven dynamics to decouple an inner companion planet from the orbital plane of a hot Jupiter.} 

{In contrast, observed adjacent exterior companions to hot Jupiters provide strong constraints on the dynamical history of a system. Our results indicate that for some regimes of parameter space, stellar spin-down can lead outer companions to attain distinct orbital planes from their adjacent hot Jupiter companions. This will occur when two conditions are met: first, there must be some initial stellar obliquity with respect to the plane containing the planets; second, the system must have assembled early enough that the planets were in roughly their observed orbital locations when the star's $J_2$ attained the value that would have caused the onset of a secular resonance. This latter condition means that for systems where non-transiting adjacent exterior companions are present, it would be possible to constrain the arrival time of the planets to their observed orbits \citep{Spalding2022}. Generally, the existence of adjacent non-transiting outer companions to hot Jupiters is consistent with in situ migration \citep{Batygin2016}, where such timing considerations are less important. }

{Our results also indicate that the presence of an inner companion to a hot Jupiter is consistent with a large set of initial conditions for both in situ and disk migration. Outer companions, however, provide more useful constraints on the arrival timing of the planetary system into its observed configuration, as secular resonances readily perturb such companions out of a mutually transiting plane. When an outer planet is also observed to transit this indicates that the outer planet was not dynamically perturbed to the extent that it would be pushed out of the transiting plane or expelled from the system. This could provide either an upper limit on the stellar obliquity, or an arrival time constraint. As discussed in \cite{Weiss2017}, the geometry of WASP-47, which has transiting inner and outer adjacent companions, provides evidence that the formation pathway is consistent with a two-stage process.}

{If hot Jupiter systems were observed to have non-transiting adjacent companions, their properties could be used to infer the value of stellar $J_2$ that the system had assembled by. Ongoing work using phase curves \citep{Lillo2021} to identify non-transiting companions in hot Jupiter systems \citep{Millholland2017, Cullen2024} could provide individual systems where such constraints could be made. }

\subsection{The Application of Our Results to the Warm Jupiter Population}
The dynamics considered in this paper requires planets to be close enough to the star that precession due to the stellar quadrupole moment $J_2$ will matter. This precession rate is given by \citep{Li2020}: 
\begin{equation}
\frac{d\omega}{dt} = \frac{3}{2} \sqrt{\frac{GM}{a^3}}  J_2 \left(\frac{R_*}{a}\right)^2 \propto a^{-5/2},
\end{equation}
which has a steep dependence on the semi-major axis $a$. As a result, planets at larger orbital distances will not experience the same degree of inclination oscillation excitation described in this work. 
Because of this, warm Jupiter systems are likely to have different dynamical regimes than those reported in this work. 
Additionally,  warm Jupiters and hot Jupiters may follow different formation and evolution pathways \citep[i.e.][]{Petrovich2016} and may experience different dynamics in general, including in their expected distributions of stellar obliquities \citep{Morgan2024} and occurrence rates of nearby companions \citep{Huang2016}. 

\subsection{Limitations and Prospects of Observations}
One significant limitation in the sample considered for this work is observational bias. First, many of the systems analyzed were discovered by missions other than Kepler, which means their completeness corrections are not well constrained. This introduces uncertainty in our understanding of the true distribution of planet properties within this class of system. 
As a general example, transit probability decreases for planets at longer orbital distances, making it more likely for outer planets to be missed, even if they are in a roughly coplanar configuration, an effect which can be compounded when non-continuous observational baselines (like those used by TESS) are considered. 
Finally, the photometric precision of the light curves imposes a detection threshold on planet radii, which depends on star-specific parameters such as stellar mass, star spot coverage, and stellar activity; planets smaller than this threshold would remain undetected regardless of their location in the system. 
Consequently, it is plausible that the systems considered in this study may host additional companions that either do not transit or are too small to be discovered with current observational capabilities.

For this reason, in this work we do not attempt a comprehensive statistical analysis of the sample. Instead, we focus on presenting the motivating dynamics that explain why outer companions may be missed, particularly in systems with non-zero stellar obliquity relative to the planet of the transiting hot Jupiter. Future completeness analyses will be needed to identify which specific systems in our sample are most likely to host unseen companions, whether non-transiting or transiting.

Finally, while the census of discovered hot Jupiters is quite large (numbering several hundred), the population of known hot Jupiters with nearby planetary companions is much smaller \citep{Hord2021}. 
Each individual system that has been discovered thus provides evidence of a new type of potential geometry. 
For example, the most recently published system, TOI-1408 \citep{Korth_2024}, presented the first confirmed case of a resonant system containing a hot Jupiter and a companion planet. Previously, none of the other known similar systems showed complete evidence of true resonance\footnote{Two different analyses of TOI-1130 suggest it may \citep{Korth2023} or may not \citep{Borsato2024} be in true resonance. Such a determination is very sensitive to fitted planet parameters, so it is not unusual for different fit methods or data sets to get different results \citep[for example][]{MacDonald2021, Weisserman2023}.}.  
In our simulations shown in Figure \ref{fig:TOI2000}, the `test' outer planet injected into the TOI-1408 system at what would have been the 3:1 mean motion commensurability went dynamically unstable, while no other planets in other trials did. 
It is possible that other similar discoveries in the future with extended data baselines \citep[i.e.,][]{Harre2024} or improved methods \citep[i.e.,][]{Chen2024} will further add complexity to this parameter space, potentially including new dynamical classes that do not match the archetypes considered in this work.

\section{Conclusion}
\label{sec:conclude}
We find that there is a rich phase space of evolutionary outcomes possible for systems containing hot Jupiters and nearby companions. In some of these dynamical modes, nearby outer companions to hot Jupiters may leave the transiting plane and be undetectable. 
In particular, we find that for a three planet system containing a hot Jupiter with nearby interior and exterior companion planets, the transit probability of the inner planet cannot be significantly affected by the combined dynamical effect of the stellar obliquity and decaying stellar quadrupole moment, while the transit probability of the outer planet can be. To first order, this is because the outer planet-hot Jupiter pair encounters a secular resonance with the stellar $J_2$ moment, while the inner planet-hot Jupiter pair does not encounter such a resonance.  
However, these results do not necessarily extend to warm Jupiter systems.

We also find that the effect of stellar obliquity is to enhance the probability of an outer companion to a hot Jupiter attaining a non-mutually transiting configuration. 
The observational impact of our results is that in systems with transiting hot Jupiters and with higher stellar obliquity, we will miss companion outer planets in transit surveys more often due to them being non-transiting as compared to in systems with lower stellar obliquity.

\begin{table*}[h!]
    \centering
    \scriptsize
    \caption{Parameters of Confirmed Hot Jupiter Systems with Nearby Companions. Parameter values for Kepler-730 come from \citet{Canas2019}; values for TOI-1130 come from \citet{Huang2020}; values for TOI-2000 come from \citet{Sha2023}, values for WASP-132 come from \citet{Hord2022}; values for WASP-47 come from \citet{Becker2015, 2016Almenara, Weiss2017, 2017Vanderburg, 2023Nascimbeni}; values for WASP-84 come from \citet{Maciejewski2023}; values for TOI-1408 come from \citet{Korth_2024}.}
    \label{tab:hot_jupiter_systems}
    \begin{tabular}{|cccccccc|}
        \hline
        \textbf{Host System} & \textbf{Planet} & \textbf{Stellar } & \textbf{Stellar } & \textbf{Orbital } & \textbf{Planet } & \textbf{Planet } &
        \textbf{Transit } \\
         &  & \textbf{Mass} & \textbf{Radius} & \textbf{Period} & \textbf{Radius} & \textbf{Mass} &
        \textbf{Midpoint} \\
       
        &  & ($M_{\Sun}$) & ($R_{\Sun}$) & (Days) & ($R_{\Earth}$) & ($M_{Jup}$) & T$_C$ (Days) \\
        \hline
        Kepler-730 & b & 1.05 & 1.41 & 6.491683 & 12.33 & 1.\footnote{Mass calculated using relation from \cite{Chen_2016} for Jovian worlds. We choose to approximate each unknown hot Jupiter mass as $\sim1 M_{jup}$ due to uncertainty in exact mass prescription.} & 2455007.633553\\
         & c & 1.05 & 1.41 & 2.851838 & 1.569 & 0.0129\footnote{Mass calculated using Mass-Radius relation for planets of radii 1.5-4 $R_{\Earth}$ \citep{2014Weiss}} & 2454965.1455 \\
        \hline
        TOI-1130 & c & 0.68 & 0.69 & 4.066499 & 3.65 & 0.0282$^b$ & 2458657.90461 \\
         & b & 0.68 & 0.69 & 8.350381 & 16.814 & 0.974 & 2458658.74627 \\
        \hline
        TOI-2000 & c & 1.08 & 1.13 & 3.09893 & 2.7 & 0.0347 & 2459110.06588 \\
         & b & 1.08 & 1.13 & 9.127055 & 8.14 & 0.257 & 2458855.2442\\
        \hline
        WASP-132 & b & 0.78 & 0.75 & 7.133514 & 10.05 &  1.$^a$ & 2459337.6080 \\
         & c & 0.78 & 0.75 & 1.011534 & 1.85 & 0.11752 & 2458597.5762 \\
        \hline
        WASP-47 & b & 1.04 & 1.14 & 4.151949 & 12.64 & 1.14401 & 2456982.97823 \\
         & c & 1.04 & 1.14 & 588.8 & N/A & 1.25508\footnote{WASP-47 c was discovered with radial velocities \citep{NeveuVanMalle2016}, and so has no measured radius and the reported mass in this table is actually $m\sin{i}$, which serves as the lower limit on the true mass.} & 2457763.1 \\
         & d & 1.04 & 1.14 & 9.030355 & 3.567 & 0.04468 & 2456988.375\\
         & e & 1.04 & 1.14 & 0.789593 & 1.808 & 0.0213 & 2456979.76455 \\
        \hline
        WASP-84 & b & 0.85 & 0.77 & 8.523496 & 10.72 & 0.692 & 2457956.71194 \\
         & c & 0.85 & 0.77 & 1.446885 & 1.95 & 0.048 & 2457952.4543\\
        \hline
        TOI-1408 & b & 1.312 & 1.53 & 2.1664 & 2.22 & 0.023 &  2458740.85 \\
         & c & 1.312 & 1.53 & 4.42587 & 25 & 1.87 & 2458739.84 \\
        \hline
    \end{tabular}
\end{table*}

\medskip
\textbf{Acknowledgments.}  
J.C.B.~has been supported by the Heising-Simons \textit{51 Pegasi b} postdoctoral fellowship. 
This research has been supported by Caltech Geological and Planetary Sciences (GPS) computing and funding, as well as essential computing resources for simulations from Konstantin Batygin. {We also thank the Caltech Housner Fund for providing funding.} This research has made use of NASA’s Astrophysics Data System.

{Software: Rebound and Reboundx, Jupyter, pandas \citep{reback2020pandas, mckinney-proc-scipy-2010}, matplotlib \citep{Hunter:2007}, numpy \citep{harris2020array}.}

\bibliography{refs.bib}

\end{document}